\documentclass[amsmath, amssymb, amsfonts, twocolumn, reprint]{revtex4}
\usepackage[german,american,english]{babel}
\usepackage{graphicx}
\usepackage{graphics}
\usepackage{dcolumn}
\usepackage{bm}
\usepackage{amssymb}
\usepackage{amsmath}
\usepackage{amsfonts}
\usepackage{epsfig}
\newcommand {\bkt} [1] {\langle #1 \rangle}

\newcommand {\td} [2] {\frac{d #1}{d #2}}

 \newcommand {\beq}{\begin{equation}}
\newcommand {\eeq}{\end{equation}}
\begin{document}
\title{Charge noise, spin-orbit coupling and dephasing of single-spin qubits}
\author{Adam Bermeister}
\affiliation{School of Physics, The University of New South Wales, Sydney 2052, Australia}
\author{Daniel Keith}
\affiliation{School of Physics, The University of New South Wales, Sydney 2052, Australia}
\author{Dimitrie Culcer}
\affiliation{School of Physics, The University of New South Wales, Sydney 2052, Australia}
\begin{abstract}
Quantum dot quantum computing architectures rely on systems in which inversion symmetry is broken, and spin-orbit coupling is present, causing even single-spin qubits to be susceptible to charge noise. We derive an effective Hamiltonian for the combined action of noise and spin-orbit coupling on a single-spin qubit, identify the mechanisms behind dephasing, and estimate the free induction decay dephasing times $T_2^*$ for common materials such as Si and GaAs. Dephasing is driven by noise matrix elements that cause relative fluctuations between orbital levels, which are dominated by screened whole charge defects and unscreened dipole defects in the substrate. Dephasing times $T_2^*$ differ markedly between materials, and can be enhanced by increasing gate fields, choosing materials with weak spin-orbit, making dots narrower, or using accumulation dots. 
\end{abstract}
\date{\today}
\maketitle

Developments in quantum computing hold considerable promise in the progress of modern information processing, and this has spurred a large experimental and theoretical effort investigating two-level systems that can be used as quantum bits (qubits). The need for scalability and long coherence times has led naturally to solid state spin-based devices, such as quantum dot spin systems, as ideal candidates for scalable qubits. The focus has been on single-spin \cite{Loss_PRA98} and singlet-triplet qubits. \cite{Petta_Science05} While GaAs quantum dots have been studied for many years, a substantial effort is also underway researching Si spin quantum computing architectures, \cite{Morton_Si_QC_QmLim_Nat11, Zwanenburg_SiQmEl_RMP13, Hao_SiDQD_SpinVlly_NC14} motivated by their compatibility with Si microelectronics and long coherence times. \cite{Feher_PR59, Abe_PRB04, Tahan_PRB05, Tyryshkin_JPC06, Wang_SiQD_ST_Relax_PRB10, Raith_SiQD_1e_SpinRelax_PRB11, Muhonen_Store_14} Recently, much effort has also been devoted to quantum dot systems with spin-orbit interactions, \cite{Friesen_Spin_Readout_PRL04, Nadj_Nat10, Palyi_SO_Res_PRL12} where spin manipulation could in principle be achieved entirely by electrical means. \cite{Bulaev_PRL07, Szumniak_PRL12, Budich_PRB12}

The coherence of a solid-state spin qubit is quantified by the relaxation time $T_1$ and the dephasing time $T_2^*$, both of which are determined by mechanisms that couple up spins with down spins. This coupling can either be direct, through the hyperfine interaction \cite{Merkulov_PRB02, Khaetskii_PRB03, Deng_PRB06, Coish_Nuclear_PSSB09} and fluctuations in the $g$-factor, \cite{Ivchenko_QW_g_fluct_SSC97} or indirect, through the joint effect of hyperfine or spin-orbit coupling and fluctuating electric fields, such as those due to phonons \cite{Erlingsson_2002_hyperfine, Golovach_PhnDecay_PRL04, Bulaev_PRL05, Schoen_Dephasing_QD_PRL06,  Prada_PRB08, Xdhu_2011_PRB, Climente_NJP13} or charge noise. \cite{Fleetwood, Jung_APL04, Huang_Charge_Noise_PRB14} Inversion symmetry breaking near an interface makes spin-orbit coupling unavoidable, even in materials such as Si in which it is weak. \cite{Wilamowski_Si/SiGeQW_Rashba_PRB02} 

Hyperfine effects typically occur on long time scales, the nuclear bath is relatively well known and can be controlled through feedback mechanisms \cite{Yacoby_Nuclear_Bath_PRL10} while in materials such as Si hyperfine coupling can be eliminated altogether through isotopic purification. \cite{Witzel_AHF_PRB07, Itoh_MRS} The spin relaxation rate due to phonons is proportional to the fifth power of the magnetic field in zinc-blende materials, in which piezoelectric electron-phonon coupling is often dominant, and to the seventh power of the magnetic field in Si, in which there is no piezoelectric coupling.\cite{Prada_PRB08} Hence phonon effects become less pronounced at low magnetic fields. They also become weaker at low temperatures. \cite{Golovach_PhnDecay_PRL04}

Noise is a well-known source of dephasing in \textit{charge} qubits. \cite{Petersson_PRL10, Dupont_SiPx2_PRL13, Dial_ST_PRL13, Paladino_1/f_RMP14} Experiments on quantum dots and point contacts have shown noise to be strong even at dilution refrigerator temperatures. \cite{Paladino_1/f_RMP14, Petersson_PRL10, Dupont_SiPx2_PRL13, Dial_ST_PRL13, Takeda_APL13, Ribeiro_PRL13, Buizert_PRL08, Hitachi_APL13, Muller_PRL06} Noise sources include $P_b$ centers, which may act as traps that charge and discharge, and tunneling two-level systems, which can be modeled as fluctuating charge dipoles. \cite{Fleetwood, Sze, Zimmermann_PRL81, Jang_JES82, Reinisch_JPCB06, Biswas_PRL99, Zimmerman_LongTermCOD_JAP08} Noise and spin-orbit coupling give rise to nontrivial physics in 2D and 1D structures. \cite{Glazov_2DEG_SOC_Dis_PhysE10, Glazov_Nwr_SpinNoise_PRL11, Sinitsyn_SpinNoiseSpect_PRL13} In quantum dot spin qubits, Ref.~\onlinecite{Huang_Charge_Noise_PRB14} has already shown that spin-orbit and noise lead to \textit{spin} relaxation, and that noise and phonon effects in general become comparable at low-enough magnetic fields. Hence, at dilution refrigerator temperatures the interplay of spin-orbit and noise may set the defining bound on spin qubit coherence. 

In this paper we build on previous decoherence work \cite{Sousa_PRB03, Hu_PRL06, Burkard_Decoh_AdvPhys08, Culcer_APL09, Sousa_BookChapter_09, RamonHu_DQD_Decoh_PRB10, Culcer_APL13} and devise a theory of dephasing due to the combined effect of charge noise and spin-orbit interactions, with two aims in mind. The first is to understand conceptually how spin-orbit and noise cause dephasing. For example, noise can give relative fluctuations between levels, virtual transitions between levels, as well as fluctuations in spin-orbit constants. We wish to isolate the terms that are responsible for dephasing. The second aim is to study the sensitivity to spin-orbit coupling across common materials with similar noise profiles. We study a sample qubit with the same specifications in different materials, we determine sample $T_2^*$s due to common noise sources, discuss the variation in $T_2^*$ across materials, and seek methods to improve $T_2^*$ generally.

\begin{figure}[tbp]
\centering
\includegraphics[width=\columnwidth]{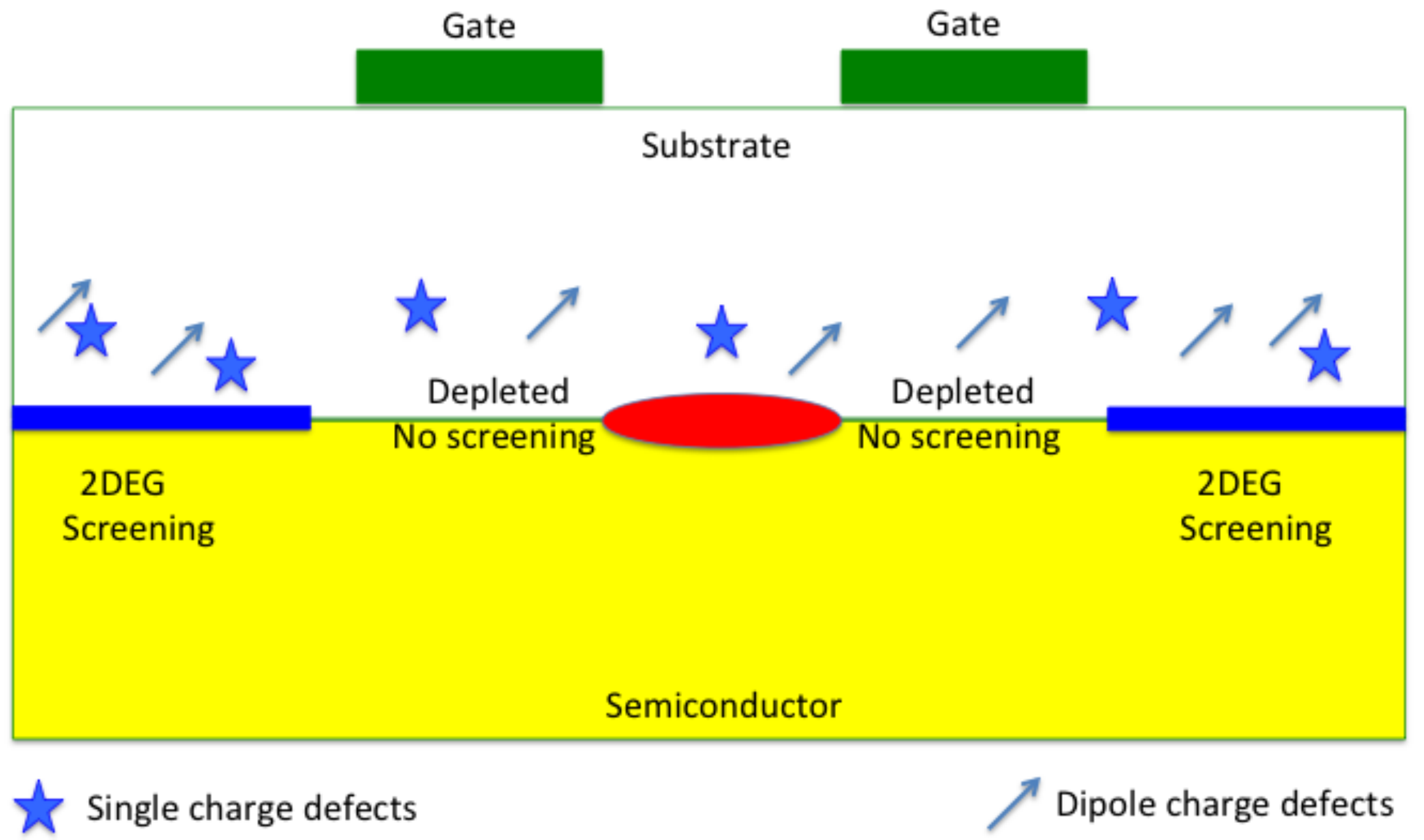}
\caption{\label{DS}
Defect locations with respect to the gate-defined quantum dot projected onto the $xz$-plane, with $\hat{\bm z}$ normal to the interface. In general a top gate is also present (not shown). The red area represents the region of the quantum dot. 
}
\end{figure}

We consider a single-spin qubit implemented in a symmetric, gate-defined quantum dot, located at a sharp flat interface (Fig.~\ref{DS}) in a dilution refrigerator at 100mK. The qubit is described by the Hamiltonian $H = H_{QD} + H_{Z} + H_{SO} + H_{N}$. The kinetic energy and confinement term
\begin{equation}
H_{QD} = - \frac{\hbar^2}{2m^*}\bigg( \frac{\partial^2}{\partial x^2} + \frac{\partial^2}{\partial y^2} \bigg) + \,  \frac{\hbar^2}{2m^*a^4}(x^2 + y^2),
\end{equation}
where $a$ is the effective dot radius and $m^*$ the effective mass. The eigenstates of $H_{QD}$ are the Fock-Darwin states, with the ground and first excited states given by
\begin{equation}
\begin{array}{rl}
\Phi_0 (x,y) &=\frac{1}{a\sqrt{\pi}} \, e^{-\left(\frac{x^2+y^2}{2a^2}\right)}\\[2\jot]
\Phi_{\pm} (x, y) &=\frac{1}{a^2\sqrt{\pi}} \, (x \pm iy) \, e^{-\left(\frac{x^2+y^2}{2a^2}\right)}.
\end{array}
\end{equation}
These have energies $\varepsilon_0=\hbar^2/2m^*a^2$ for the orbital ground state and $\varepsilon_{1}=3\hbar^2/2m^*a^2$ for the twofold degenerate first orbital excited state. The orbital level splitting is assumed to be the dominant scale, so that only the ground and first excited states are considered. The Zeeman Hamiltonian $H_Z = \frac{1}{2}g\mu_B {\bm \sigma} \cdot {\bm B}$, with ${\bm \sigma}$ the vector of Pauli spin matrices. Since ${\bm B}$ is constant, the orbital effect of ${\bm B}$ can be absorbed into the effective dot radius $a$. We have also not taken into account multi valley effects in Si. For a certain interaction to couple valley states appreciably, it must be sufficiently sharp in real space. Neither the spin-orbit coupling due to the interface field nor the electric field of the defect satisfy this requirement -- even though these interactions are important in relaxation in particular around hot spots. \cite{Peihao_SpinValley_14}

The spin-orbit term $H_{SO} = H_{R1} + H_{D1}$. The Rashba term $H_{R1} = \alpha(t) \, {\bm \sigma}\cdot ({\bm{k}}\times \hat{\bm z})$, stems from structure inversion asymmetry, where ${\bm k} = - i{\bm \nabla}$ here is an operator in real space, $\hat{\bm z}$ is the unit vector perpendicular to the interface, and $\alpha$ is determined by a material specific parameter as well as the interface electric field $E_z$. \cite{Winkler2003} Thus $\alpha$ is also sensitive to stray electric fields and fluctuates in time, thus we let $\alpha(t) = \alpha \, [1+\lambda(t)]$ where $\lambda \ll 1$. For a quantum dot on a (001) surface the linear Dresselhaus term $H_{D1} = \beta (\sigma_y k_y - \sigma_x k_x)$ is usually the dominant bulk inversion asymmetry contribution,\cite{Winkler2003} where $\beta = \beta_3 \, (\pi/w)^2$, with $\beta_3$ a material-specific parameter and $w$ the width of the $z$-confinement perpendicular to the interface. Since $H_{D1}$ can be obtained from $H_{R1}$ by a $\pi/2$ spin rotation, they give rise to qualitatively similar physics. In Si $\beta = 0$ due to inversion symmetry, whereas Rashba spin-orbit coupling is expected generally in a 2D electron gas near an interface, and should be present in all gate-defined dots. In zincblende structures $H_{R1}$ and $H_{D1}$ may comparable in magnitude in certain parameter regimes, though for $E_z \approx 10^7$ Vm$^{-1}$ ($\equiv$ 0.1 V/10 nm), the Rashba term is expected to be the dominant spin-orbit contribution.

The noise Hamiltonian $H_N(t)$ is a random function of time. We do not include gate noise in our model, and we first consider random telegraph noise (RTN). In the simplest case, in which the qubit is only sensitive to one defect, $H_N$ represents a fluctuating Coulomb potential, screened by the nearby 2D electron gas. The 2D screened Coulomb potential $U_{scr}$ is written in terms of its Fourier transform, which is a function of momentum ${\bm q}$ \cite{Davies}
\begin{equation}
U_{scr}(r) = \frac{e^2}{2\epsilon_0\epsilon_r} \int_0^{2k_F} \frac{d^2 q}{(2\pi)^2} \frac{e^{-i\bm{q}\cdot \bm{r}}}{q+q_{TF}},
\end{equation}
with $\epsilon_r$ the relative permittivity, $q_{TF}$ the Tomas-Fermi wave vector, and $k_F$ the Fermi wave vector (the contribution from $q > 2k_F$ is negligible \cite{Culcer_APL13}). The matrix elements entering $H_N$ are $v_0 = \langle \Phi_0 |U_{scr}| \Phi_0 \rangle$, $v_1 = \langle \Phi_{\pm}| U_{scr}| \Phi_{\pm} \rangle$, $v_2 = \langle \Phi_0|U_{scr}|\Phi_{\pm} \rangle \approx \langle \Phi_{\pm}|U_{scr}|\Phi_0 \rangle$ and $v_3 = \langle \Phi_{\pm}|U_{scr}|\Phi_{\mp} \rangle \approx \langle \Phi_{\mp}|U_{scr}|\Phi_{\pm} \rangle$. For RTN we can write $v_i(t) = v_i (-1)^{N(t)}$ for $i = 0,1, 2$, and $N(t)=0,\ 1$ is a Poisson random variable with switching time $\tau$. \footnote{The effect of fluctuators with $\tau \ge 1$ $\mu$s can be eliminated in experiment through dynamical decoupling. Hence, on physical grounds, we impose 1 $\mu$s as a cutoff for the switching time.} 

Additional (extrinsic) spin-orbit coupling arises from the electric field of the defect itself. Yet for a charge defect located 40 nm away from the dot this field is several orders of magnitude smaller than the interface electric field $E_z$. Because the matrix element involved is second order in $v_i$, the contribution this makes to dephasing is many orders of magnitude smaller than the Rashba interaction due to $E_z$, and will not be considered further.

In the basis $\{\Phi_{0\uparrow}, \Phi_{0\downarrow}, \Phi_{+\uparrow}, \Phi_{+\downarrow}, \Phi_{-\uparrow}, \Phi_{-\downarrow}\}$, with $\uparrow, \downarrow$ representing up and down spins, the Hamiltonian reads
\begin{equation}\label{H}
\begin{split}
H = \left(\begin{array}{cc|cccc}
\varepsilon_0^+ & 0 & v_2 & s_R & v_2 & is_D \\[2\jot]
0 & \varepsilon_0^- & is_D & v_2 & -s_R & v_2 \\[1\jot]\hline
\rule{0pt}{1\normalbaselineskip} v_2 & -is_D & \varepsilon_1^+ & 0 & v_3 & 0 \\[2\jot]
s_R & v_2 & 0 & \varepsilon_1^- & 0 & v_3 \\[2\jot]
v_2 & -s_R & v_3 & 0 & \varepsilon_1^+ & 0 \\[2\jot]
-is_D & v_2 & 0 & v_3 & 0 & \varepsilon_1^-
\end{array}\right)
\end{split}
\end{equation}
where $\varepsilon_0^{\pm}(t) = \varepsilon_0 + v_0(t) \pm \varepsilon_Z$ and $\varepsilon_1^{\pm}(t) = \varepsilon_1 + v_1 (t) \pm \varepsilon_Z$ are the Zeeman-split orbital levels including the noise terms, the Zeeman energy $\varepsilon_Z=\frac{1}{2}g\mu_B B$, and the spin-orbit terms $s_D=\beta/a$ and $s_R(t) = s_R\, [1 + \lambda(t)]$, with $s_R=\alpha/a$ (not a function of time). 

The qubit subspace is simply the Zeeman-split orbital ground state $\{\Phi_{0\uparrow}, \Phi_{0\downarrow}\}$, which has been singled out in the top left hand corner of Eq.\ \ref{H}. These two states are coupled by $H_N$ to spin-aligned orbital excited states and by $H_{SO}$ to orbital excited states with anti-aligned spin. By projecting $H$ onto this subspace we encapsulate the combined effect of spin-orbit coupling and noise in an \textit{effective} qubit Hamiltonian $H_{qbt}$. To achieve this, we carry out a Schrieffer-Wolff transformation, eliminating higher orbital excited states. \cite{Winkler2003, Aleiner_PRL01, Golovach_PhnDecay_PRL04, Stano_PRL06} Keeping terms up to the second order in this transformation,
\begin{equation}\label{Hqbt}
H_{qbt}(t) = H_Z - \frac{2  \varepsilon_Z\{v_2(t)[s_R(t) \sigma_x + s_D \sigma_y] + [s_R(t)^2+s_D^2]\sigma_z\}}{[\delta \varepsilon + \delta v(t)]^2}
\end{equation}
where $\delta \varepsilon = \varepsilon_0 - \varepsilon_1$ (not a function of time) and $\delta v(t) =v_0(t) - v_1(t)$. We retain only terms of first order in ${\varepsilon_Z}$ and $\delta v$. Equation (\ref{Hqbt}) implies that, in addition to $H_Z$, there exists an effective Zeeman term $\frac{1}{2}\bm{\sigma}\cdot \bm{V}(t)$, where $\bm{V}(t)$ represents an effective fluctuating effective magnetic field due to the combined action of spin-orbit and noise. For convenience $\bm{V}$ has units of energy and, for RTN, $\bm{V}(t) = {\bm V} (-1)^{N(t)}$. We will also use $V(t) = |\bm{V}(t)|$ for the magnitude of ${\bm V}$. Since the Rashba and Dresselhaus contributions are added in quadrature, there is no sweet spot for dephasing. 

The noise matrix elements appearing in $H_{qbt}$ may be divided into two categories. The diagonal elements $v_0(t), v_1(t)$ cause different orbital levels to fluctuate by different amounts, while the off-diagonal element $v_2(t)$ causes transitions between different orbital levels. If the qubit is initialized in an off-diagonal state, the diagonal elements ($\sigma_z$) in $H_{qbt}$ give dephasing. These terms involve the intraband matrix elements $v_0(t), v_1(t)$ of the defect potentials. An additional contribution comes from fluctuations in $\alpha$, which lead to fluctuations in $s_R$ itself. These fluctuations can be interpreted as a modulation of the $g$-factor, and are expected to come from defects in the substrate right above the dot, which modify $E_z$. Since the dot region is depleted, whole charge defects cannot fluctuate, except in the very special case in which the defect lies right above the dot. Hence defects contributing to $E_z$ are expected to be mostly charge dipoles, stemming for example from passivated traps. Although these are weaker than whole charge defects, they are unscreened, leading to a subtle competition. Thus, generally, dephasing stems from noise matrix elements that cause relative fluctuations between orbital levels. In contrast, if the qubit is initialized in the spin-up state, the off-diagonal elements ($\sigma_x$) in $H_{qbt}$ give relaxation ($T_1$ processes), which was studied in detail in Ref.~\onlinecite{Huang_Charge_Noise_PRB14}. These elements are of first order in $\alpha$ and involve the interband defect matrix element $v_2(t)$. \footnote{We expect whole charge defect potentials to be dominant in relaxation since they are much stronger than dipole potentials \cite{Culcer_APL13}.}

In order to study dephasing further and obtain quantitative estimates of $T_2^*$, we focus on a single-spin qubit described by a spin density matrix $\rho(t)$. The spin density matrix satisfies the quantum Liouville equation
\begin{equation}\label{eq:QLE}
\td{\rho}{t} + \frac{i}{\hbar} \, [H_{qbt}, \rho] = 0.
\end{equation}
The spin density matrix $\rho(t) = \frac{1}{2}\bm{\sigma}\cdot \bm{S}(t)$. Any spin component $S_i$ can be found as $S_i(t) = \text{tr} \, [\sigma_i\rho(t)]$, with tr the matrix trace. We restrict our attention to RTN for the time being. Using the time evolution operator $e^{-(i/\hbar)\int^t_0H_{qbt}(t')\ dt'}$, we obtain the general time evolution of the spin as
\begin{equation}
{\bm S} (t) = {\bm S}_{0} \cos h - ({\bm S}_{0} \times \hat{\bm h}) \sin h + \hat{\bm h} (\hat{\bm h} \cdot {\bm S}_0) (1 - \cos h),
\end{equation}
where we have defined ${\bm S}_0 \equiv {\bm S}(t=0)$ and, for RTN, ${\bm h} (t)=({\bm V}/\hbar)\int^t_0 (-1)^{N(t')}\ dt'$, with $h(t) = |{\bm h} (t)|$. The two components of ${\bm h}$ have exactly the same time evolution. Since $|{\bm B}| \gg |{\bm V}$, if ${\bm S}_0 = S_{0x} \hat{\bm x}$ is initialised, $S_x(t) \approx S_{0x} \, \cos{[h(t)]}$. Averaging over noise realisations \cite{Sousa_PRB03, Culcer_APL09, Culcer_APL13} 
\begin{equation}
\langle \cos{[h(t)]} \rangle \rangle=e^{-t/\tau}\bigg(\frac{\sinh{\Xi t}}{\Xi \tau}+\cosh{\Xi t}\bigg),
\end{equation}
where $\Xi=\sqrt{(\hbar/\tau)^2-V^2}/\hbar$. All systems of interest in this work satisfy $V^2\ll(\hbar/\tau)^2$, in which case we may approximate $\sqrt{\left(\frac{\hbar}{\tau}\right)^2-V^2}\approx\frac{\hbar}{\tau}\left(1-\frac{V^2\tau^2}{2\hbar^2}\right)$. When the denominator of the $\sinh$ is expanded in $(V\tau/\hbar)^2$, only the leading term in the expansion may be retained. Physically, in this case, the time dynamics of $h(t)$ are a random walk in time, and the spread in $\cos h(t)$ leads to motional narrowing. As a result, the initial spin decays exponentially as $S_x(t) \approx S_{0x} \, e^{-t/T^*_2}$, where 
\begin{equation}
\bigg(\frac{1}{T_2^*}\bigg)_{RTN} = \frac{V^2\tau}{2\hbar^2}.
\end{equation} 
For whole charge defects, where dephasing is dominated by fluctuations in the orbital energy, we may set $\lambda(t)=0$ and retain $V_{wh}(t) = 8 \, (s_R^2 + s_D^2) \, \varepsilon_Z\delta v(t)/(\delta \varepsilon)^3$. For dipole charge defects we have $V_{dip}(t) = 8 \, (s_R^2 + s_D^2) \, \varepsilon_Z\lambda(t)/(\delta \varepsilon)^2$.

\begin{figure}[tbp]
\centering
\includegraphics[width = \columnwidth]{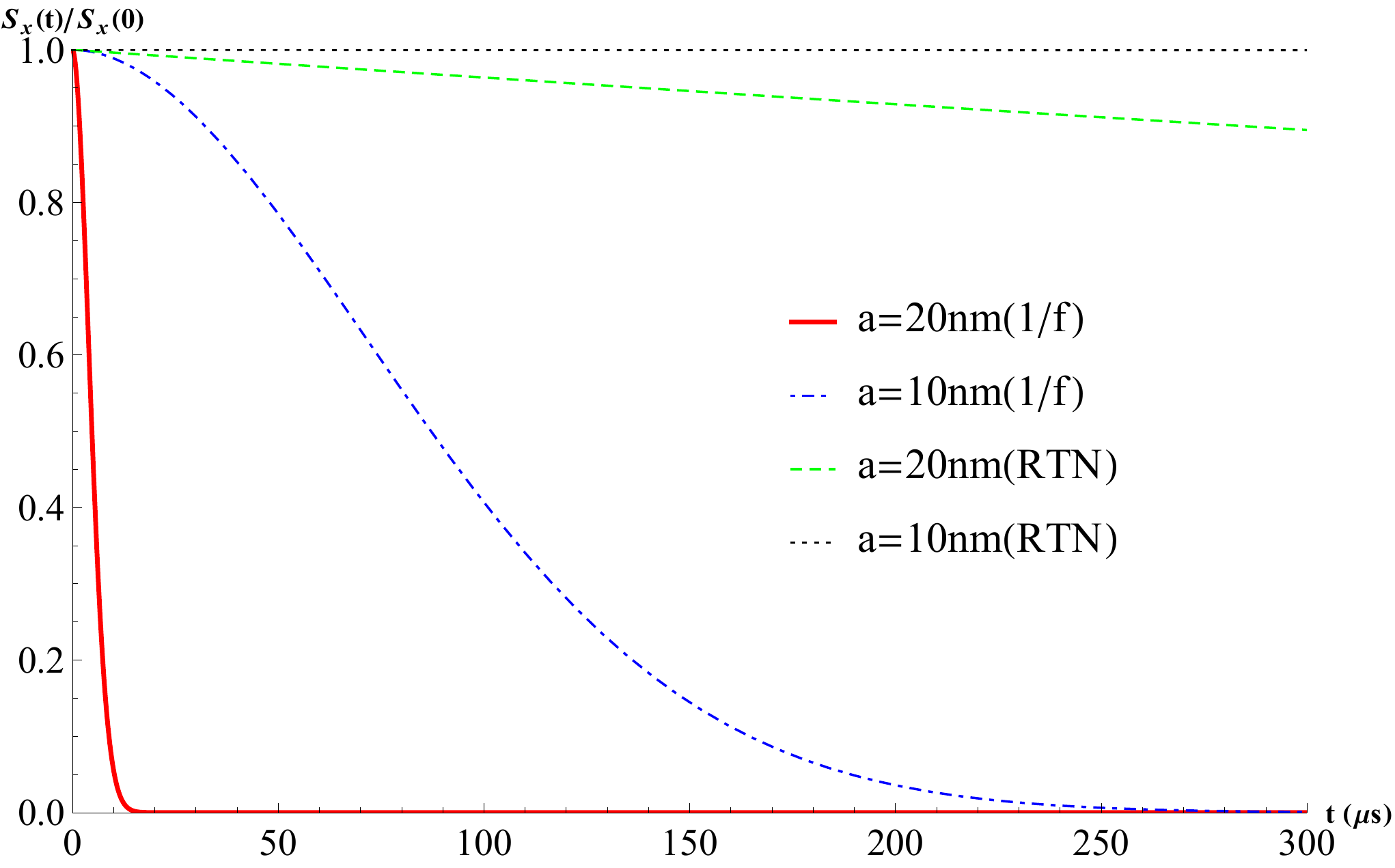}
\caption{\label{fig:1overf} Time evolution of the initated spin for different dot radii $a$ in Si/SiGe, $\omega_0$=1 s and other values as in Table \ref{tab:T2}.}
\end{figure}

We turn our attention next to $1/f$ noise. In semiconductors $1/f$ noise is Gaussian \cite{Kogan} and is fully described by its spectral density $S(t - t') = \bkt{H_N(t)H_N(t')}$. The Fourier transform of this spectral density has the form $S(\omega) = \frac{\gamma k_BT}{\omega}$, where $\gamma$ is a parameter typically inferred from experiment. Based on our estimates for RTN above we expect whole charge defects to dominate dephasing. Hence, for the effective fluctuating magnetic field ${\bm V}(t)$ acting in the qubit subspace, we may write approximately $S_V(\omega) \approx \displaystyle \bigg[\frac{8(s_R^2 + s_D^2)\varepsilon_Z}{(\delta \varepsilon)^3}\bigg]^2 S(\omega)$. To study dephasing, we write $S_x(t) = S_{0x} \, e^{-\chi(t)}$, where
\begin{equation}
\chi (t) = \frac{2\gamma k_BT}{\hbar^2}\, \bigg[\frac{8(s_R^2 + s_D^2)\varepsilon_Z}{(\delta \varepsilon)^3}\bigg]^2 \int_{\omega_0}^{\infty}d\omega \, \frac{\sin^2\omega t/2}{\omega^3}.
\end{equation}
The low-frequency cut-off $\omega_0$ is usually taken to be the inverse of the measurement time. At times $t \ll 1/\omega_0$ such as we consider here, we can approximate
\begin{equation}
\chi (t) \approx \bigg(\frac{t}{T_2^*}^2\bigg) \, \ln \frac{1}{\omega_0t},
\end{equation} 
where the dephasing time is estimated by
\begin{equation}
\bigg(\frac{1}{T_2^*}\bigg)_{1/f} \approx \sqrt{\frac{\gamma k_BT}{2\hbar^2}} \, \bigg[\frac{8(s_R^2 + s_D^2)\varepsilon_Z}{(\delta \varepsilon)^3}\bigg].
\end{equation}
Since this definition of $T_2^*$ is approximate, we plot the full time evolution of $S_x(t)$ in Fig.\ \ref{fig:1overf}.

\begin{table}[t!]
  \caption{\label{tab:T2}Sample $T_2^*$ for a quantum dot with $a = 20$ nm, $\lambda=4\times 10^{-4}$, $\tau=1\mu$s and the defect distance is 40 nm (for RTN), $E_z=20$ MV/m, $\varepsilon_Z = 60$ $\mu$eV, T = 0.1 K, $\alpha$ from Refs.~\onlinecite{Tahan_PRB05, Winkler2003}, $\beta$ from Ref.~\onlinecite{Winkler2003} and $S(\omega)$ for $1/f$ noise estimated from Refs.\ \onlinecite{Takeda_APL13, Petersson_PRL10}. Following Ref.~\onlinecite{Tahan_PRB05}, the confinement perpendicular to the interface ($\parallel \hat{\bm z}$) is represented by a square well of width 15 nm. For Si the valley splitting is assumed large.}
 $\arraycolsep 0.33em
   \begin{array}{cccccc} \hline\hline
   \ & \alpha(\text{peV\ m}) & \beta(\text{peV\ m})  & (T_2^*)^{RTN}_{wh} & (T_2^*)^{RTN}_{dip} & (T_2^*)^{1/f}_{wh} 
    \\ \hline
    \text{Si/SiGe}  & 0.02 & 0 & 3 \, \rm{ms}  & 18 \, \rm{s}                & 20 \, \rm{\mu s}  \\
    \text{GaAs} & 1.0 & 0.12 & 60 \, \rm{ns}  & 280 \, \rm{\mu s}  & 20 \, \rm{ns} \\
    \text{InAs} & 23 & 0.12 & 40 \, \rm{ps} & 65 \, \rm{ns}  & 900 \,\rm{ps} \\
    \text{InSb} & 105 & 3.4 & 1 \, \rm{ps} & 1 \, \rm{ns} & 200 \, \rm{ps} \\ \hline\hline
  \end{array}$
\end{table}

We consider a sample dot with radius $a = 20$ nm located at $x=y=0$, and $\alpha$ as calculated in Refs.~\onlinecite{Tahan_PRB05, Winkler2003}. For a defect in the plane of the dot with $x = 40$ nm, $v_0=23\ \mu$eV, $v_1=71\ \mu$eV and $v_2=31\ \mu$eV. Next we estimate the change in $\alpha$ due to a dipole defect right above the dot ($x=y=0$) and $z=3$ nm away from it. \cite{Sze, Fleetwood} The potential of an unscreened charge dipole located a distance ${\bm R}_{D}$ away from the dot, is $U_{dip} (R_{D}) = \frac{{\bm p}\cdot\hat{\bm R}_{D}}{4\pi \varepsilon_0 \varepsilon_r R_{D}^2}$. The charge dipole has dipole moment ${\bm p} = -e{\bm l}$, where ${\bm l} = (l_x, l_y, l_z)$. We take the expectation value of $U_{dip} (R_{D})$ using $\Phi_0$, and compare it with the matrix element of $e E_z z$, yielding $\lambda = 4 \times 10^{-4}$. We use this figure in all our estimates since $\varepsilon_r$ for all materials considered are of very similar magnitudes. For $1/f $ noise we extract $\gamma$ from experiment. For Si/SiGe we use Ref.~\onlinecite{Takeda_APL13}, and for GaAs Ref.\ \onlinecite{Petersson_PRL10}, while for InAs and InSb, in the absence of experimental data, we use the same $S(\omega)$ as for GaAs. 

The results are listed in Table \ref{tab:T2}, which is the central result of this paper. For all materials, whole charge defects dominate dephasing. Table \ref{tab:T2} shows that terms of second-order in spin-orbit are effective in causing dephasing, and the dependence on $\alpha^2$ causes vast differences in dephasing times $T_2^*$ between materials. Hence, using materials with a small $\alpha$ such as Si can improve coherence enormously. If spin-orbit coupling is needed for electric dipole spin resonance, increasing $E_z$ will align the charge dipoles. Although that increases $\alpha$ and with it dephasing, it also reduces the gate time by an equal amount. Moreover, for $1/f$ noise, $T_2^* \propto a^{-4}$, so by halving the dot radius the dephasing time can be increased by an order of magnitude (Fig.\ \ref{fig:1overf}; for RTN, $T_2^* \propto a^{-8}$). One can also use pulse sequences, \cite{Ribeiro_PRL13} lower the temperature to reduce $S(\omega)$, use accumulation dots, in which there is no nearby 2DEG, or focus on reducing charge noise. \cite{Buizert_PRL08, Takeda_APL13, Hitachi_APL13}

Following existing calculations of $\alpha$, \cite{Tahan_PRB05} we have taken the $\hat{\bm z}$-confinement in the form of a square well, whereas semiconductor interfaces are more accurately described by a triangular well. Nevertheless, since the form of $H_{R1}$ and $H_{D1}$ is dictated by symmetry, they will be identical in structure for triangular confinement, thus we may simply treat $\alpha$ and $\beta$ as phenomenological parameters. Finally, fluctuations in $w$ affect $\beta$. Although this effect, likewise driven by fluctuating dipoles, can be calculated in the same way as the renormalization of $\alpha_0$ by $\lambda (t)$, we expect its contribution to be minor, in exact analogy with $H_{R1}$. 

In summary, we have shown that spin-orbit coupling and charge noise are an effective source of dephasing in single-spin qubits even in materials such as GaAs in which spin-orbit coupling is weak. Based on realistic experimental parameters vast differences in spin dephasing times exist between common materials. In the future we will devise a full model of $1/f$ noise \cite{Martin_TTLS_PRB06} as an ensemble of incoherent RTNs, \cite{Muller_PRL06} where qubit dynamics is nontrivial. \cite{Burkard_NonMarkov_PRB09} Dephasing of hole spin qubits, in which spin-orbit interactions are also strong but the heavy hole-light hole coupling cannot be ignored, will likewise be studied in a future publication.

\acknowledgments

We thank R. Winkler, Sven Rogge, Joe Salfi, Andrea Morello, K. Takeda, Amir Yacoby, L. Vandersypen, Neil Zimmerman, S. Das Sarma, Alex Hamilton, Xuedong Hu, Guido Burkard, Mark Friesen, Andrew Dzurak, Menno Veldhorst, Floris Zwanenburg, J. R. Petta, and Matt House for enlightening discussions. 


%

\end{document}